\documentclass[fleqn,10pt]{wlscirep}
\usepackage[utf8]{inputenc}
\usepackage[T1]{fontenc}
\usepackage{multirow}

\title{Dissociation dynamics of the diamondoid adamantane upon photoionization by XUV femtosecond pulses}

\author[1,2,*]{Sylvain Maclot}
\author[1]{Jan Lahl}
\author[1]{Jasper Peschel}
\author[1]{Hampus Wikmark}
\author[1]{Piotr Rudawski}
\author[1]{Fabian Brunner}
\author[1]{Hélène Coudert-Alteirac}
\author[3]{Suvasthika Indrajith}
\author[3]{Bernd A. Huber}
\author[4,5,6]{Sergio D\'iaz-Tendero}
\author[7]{N\'estor F. Aguirre}
\author[3]{Patrick Rousseau}
\author[1,*]{Per Johnsson}

\affil[1]{Department of Physics, Lund University, P.O. Box 118, 22100 Lund, Sweden}
\affil[2]{Biomedical and X-Ray Physics, Department of Applied Physics, AlbaNova University Center, KTH Royal Institute of Technology, SE-10691 Stockholm, Sweden}
\affil[3]{Normandie Universit\'e, ENSICAEN, UNICAEN, CEA, CNRS, CIMAP, 14000 Caen, France}
\affil[4]{Departamento de Qu\'imica, M\'odulo 13, Universidad Aut\'onoma de Madrid, 28049 Madrid, Spain}
\affil[5]{Condensed Matter Physics Center (IFIMAC), Universidad Aut\'onoma de Madrid, 28049 Madrid, Spain}
\affil[6]{Institute for Advanced Research in Chemical Sciences (IAdChem), Universidad Aut\'onoma de Madrid, 28049 Madrid, Spain}
\affil[7]{Theoretical Division, Los Alamos National Laboratory, Los Alamos, New Mexico 87545, USA}
\affil[*]{smaclot@gmail.com and per.johnsson@fysik.lth.se}


\begin{abstract}
This work presents a photodissociation study of the diamondoid adamantane using extreme ultraviolet femtosecond pulses. The fragmentation dynamics of the dication is unraveled by the use of advanced ion and electron spectroscopy giving access to the dissociation channels as well as their energetics. To get insight into the fragmentation dynamics, we use a theoretical approach combining potential energy surface determination, statistical fragmentation methods and molecular dynamics simulations. We demonstrate that the dissociation dynamics of adamantane dications takes place in a two-step process: barrierless cage opening followed by Coulomb repulsion-driven fragmentation.   
\end{abstract}
\begin{document}
\flushbottom
\maketitle
\thispagestyle{empty}

\section{\label{sec:Intro}Introduction}

\begin{figure*}
\centering
\includegraphics[width=0.2\textwidth]{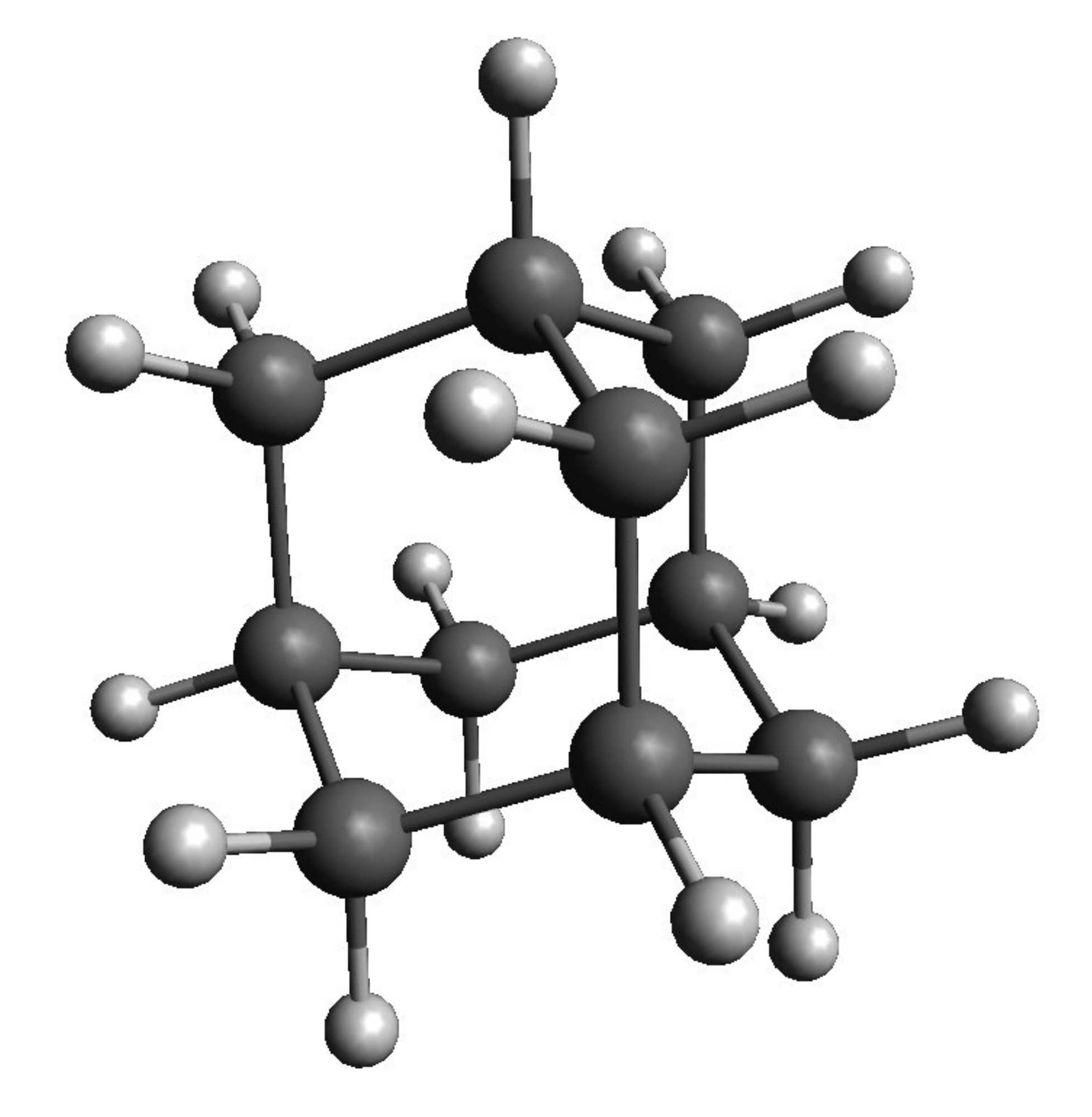}
\caption{\label{fig:struct} Structure of adamantane.}
\end{figure*}

Diamondoids are a class of carbon nanomaterials based on carbon cages with well-defined structures formed by $\rm C(sp^3)-C(sp^3)$-hybridized bonds and fully terminated by hydrogen atoms. 
All diamondoids are variants of the adamantane molecule, the most stable among all of the isomers with the formula $\rm C_{10}H_{16}$, shown in Figure \ref{fig:struct}.
On Earth, diamondoids are naturally found in petroleum deposits and natural gas reservoirs, and  their most common applications are for the characterization of petroleum and gas fields, offering possibilities to e.g. trace the source of oil spills~\cite{Schwertfeger2008}. 
Today, diamondoids are attracting increasing interest for use as an applied nanomaterial in e.g. nano- and optoelectronics as well as in biotechnology and medicine due to their high thermal stability and well-defined structure in combination with no known toxicity~\cite{Stauss2017}. 
In space, diamondoids have been found to be the most abundant component of presolar grains~\cite{Anders1993}, and due to their high stability they are thus also expected to be abundant in the interstellar medium~\cite{Henning1998}. 

However, when compared to laboratory measurements based on infrared spectroscopy~\cite{Oomens2006}, astronomical observations show a deficiency of diamondoids in the interstellar medium which to date is not completely understood~\cite{Acke2006}.
The first ionization limit in diamondoids lies around $8-9$~eV with a maximum in the ionization yield between 10 and 11~eV~\cite{Lenzke2007}, close to the hydrogen Lyman-$\alpha$ line, and the efficient production of cations followed by dissociation has been suggested as a possible explanation for the apparent lack of diamondoids in the interstellar medium. 
Steglich \textit{et al.} investigated the stability of diamondoid cations using ultraviolet irradiation, finding that rapid loss of a neutral hydrogen followed ionization~\cite{Steglich2011}. Since then, further studies have suggested that  small hydrocarbons are also created as dissociation products.
In a recent work at the Swiss Light Source, vacuum ultraviolet radiation (9-12~eV) was used in combination with threshold photoelectron and photoion coincidence detection to determine the appearance energies and branching ratios of the resulting photofragments of the singly charged adamantane cation~\cite{Candian2018}.
The study reveals, in addition to the expected hydrogen loss, dissociation via a number of parallel channels which all start with an opening of the carbon cage and hydrogen migration indicating that the low photostability of adamantane could explain its deficiency in astronomical observations. 
While this study was recently complemented by a first time-resolved study~\cite{marciniak2019electron}, to date no results have been published on the dissociation dynamics of multiply charged adamantane molecules.

In this work, we study the fragmentation dynamics of the adamantane dication after ionization by extreme ultraviolet (XUV) femtosecond pulses, the use of which ensures prompt and well-defined ionization.
The experimental technique used in this study is based on correlated ion and electron spectroscopy, enabling the characterization of the charged products of interaction (identification and energetics). 
The support of various theoretical methods such as molecular dynamics simulations, potential energy surface determination and statistical fragmentation models, helps to unravel the fragmentation dynamics of such a complex molecular system.

\section{\label{sec:Method}Methods}
\subsection{\label{sec:Experim}Experiments}

\begin{figure}[b]
\centering
\includegraphics[width=.9\textwidth]{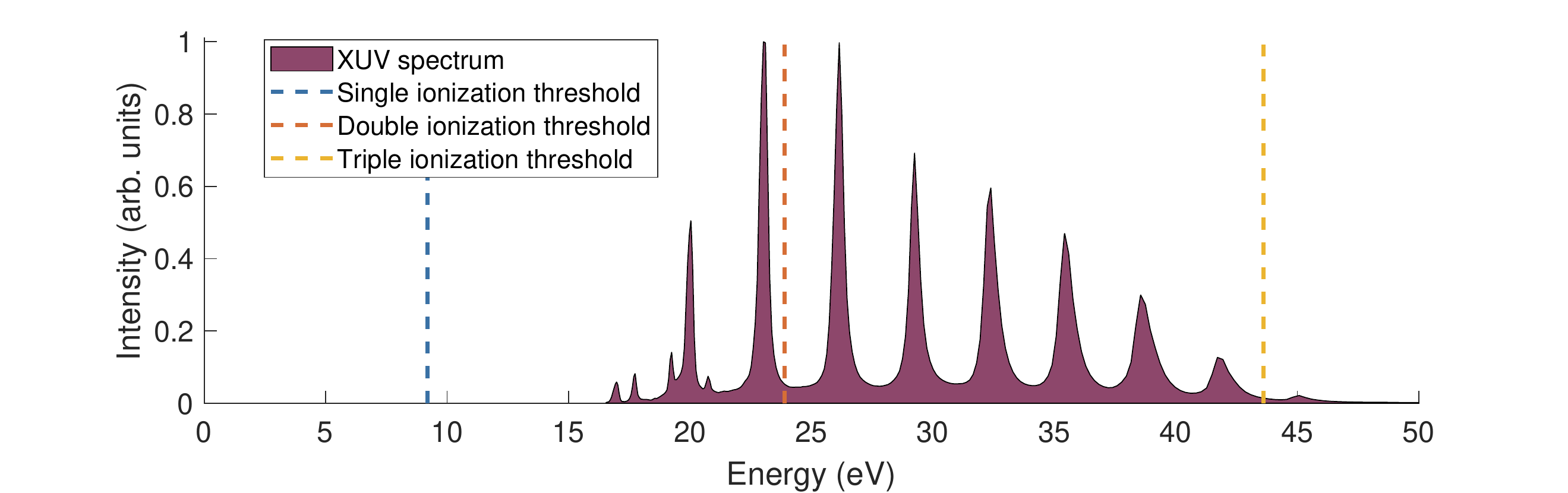}
\caption{\label{fig:XUVSpec}XUV spectrum and the first three ionization thresholds of adamantane.}
\end{figure}

The high-intensity XUV beamline at the Lund Attosecond Science Centre provides trains of attosecond pulses in the XUV spectral region using the high-order harmonic generation (HHG) technique~\cite{Ferray1988,McPherson87}. 
This is achieved by focusing an intense infrared (IR) pulse (high-power Ti:Sapphire chirped pulse amplification laser with a pulse energy of 50~mJ, a central wavelength of 810~nm, a pulse duration of 45~fs and a repetition rate of 10~Hz) into an argon gas medium (6 cm long cell) in a loose focusing geometry ($\sim 8$~m focal length)~\cite{Manschwetus2016}. 
The train of XUV attosecond pulses, with a total duration of 20~fs, contains around 15~attosecond pulses with an estimated individual pulse duration of approximately 300~as, spaced by 1.35~fs.
The photon energy spectrum of the produced XUV light is a characteristic harmonic comb spanning from $\sim 20$ to 45~eV (Figure~\ref{fig:XUVSpec}).
Then, the XUV light is micro-focused ($\rm \sim 5\times5 \, \mu m^2$) on target via a double toroidal mirror~\cite{Coudert2017} and with a pulse train energy on target of around 10~nJ this leads to intensities of the order of $10^{12}$~W$\cdot$cm$^{-2}$.

Adamantane molecules, $\rm C_{10}H_{16}$ (powder from Aldrich with $>$99\% purity), are produced in the gas phase by a pulsed Even-Lavie valve~\cite{Even2014,Even2015}, heated to 100$^{\circ}$C and using He as a carrier gas, in the form of a cold and collimated supersonic jet.

The photon-molecule interaction leads to the formation of highly excited singly and doubly charged adamantane molecules (single and double ionization thresholds $\rm IT_1 \sim 9.2$~eV and $\rm IT_2 \sim 23.9$~eV - see Figure~\ref{fig:XUVSpec}). 
The trication is only produced in negligible amounts since the triple ionization threshold is $\rm IT_3 \sim 43.6$~eV (Figure~\ref{fig:XUVSpec}) and thus is not discussed in the following. The charged products of interaction are detected by a double-sided velocity map imaging (VMI) spectrometer~\cite{Rading2018} giving access to the kinetic energy distributions of ions and electrons on a shot-to-shot basis. 
In addition, the ion side of the spectrometer can measure the time-of-flight (TOF) of the ions, providing the mass spectrum.
Despite the high count rates (several tens of counts per pulse), the use of the partial covariance technique~\cite{Frasinski2016} enables us to disentangle the contributions of the different fragmentation channels.
For instance, applying this technique on single-shot ion TOF spectra gives the possibility to produce ion-ion correlation maps that reveal the dissociation dynamics of the doubly charged adamantane molecules. 
In addition, the use of this technique on single-shot ion TOF spectra and the single-shot ion VMI data gives access to the kinetic energy distribution of specific ionic fragments.

\subsection{\label{sec:Theo}Theory}
Three different theoretical methods have been used: (i) molecular dynamics (MD) simulations in the framework of the density functional theory (DFT) and the density functional tight binding (DFTB) method, (ii) exploration of the potential energy surface (PES) employing the DFT, and (iii) statistical fragmentation using the Microcanonical Metropolis Monte-Carlo (M3C) method.

\subsubsection*{Molecular dynamics simulations}
The molecular dynamics approach, using the DFTB method\cite{Porezag1995,Seifert1996}, has been used to compute the lifetime of doubly-ionized and excited adamantane.
To this end, we have considered double ionization of adamantane in a Franck–Condon way; that is, our molecular dynamics simulations start from the optimized geometry of the neutral adamantane molecule after removal of two electrons. 
We assume that the electronic excitation energy ($\rm E_{exc}$) is rapidly redistributed into the nuclear degrees of freedom and thus we run these simulations in the electronic ground state.
We have taken four values of excitation energy corresponding to the relative energy of the highest order of harmonics in the XUV spectrum (Figure~\ref{fig:XUVSpec}) with respect to IT$_2$ ($\rm E_{exc}$ = 8.46, 11.50, 14.63 and 17.80~eV, which correspond to temperature values 2520, 3450, 4360 and 5300~K, respectively). 
The used $\rm E_{exc}$ values can be considered as the upper limits of the remaining electronic excitation energy after ionization with the four highest energy harmonics.
This excitation energy is randomly distributed into the nuclear degrees of freedom in each trajectory. Molecular dynamics trajectories are propagated up to a maximum time of $\rm t_{max}$= 1, 5, 10, 20 and 100~ps.
For each value of excitation energy and propagation time a set of 1000 independent trajectories are considered (that is, the initial conditions are separately established in each set of trajectories). Statistics are then carried out over these trajectories to obtain information on the survival time of the doubly-ionized adamantane with different excitation energies. 
To ensure adiabaticity in the simulations a time step of $\Delta$t=0.1 fs is used.
These simulations have been carried out with the deMonNano code \cite{deMonNano} and the results show that even with these relatively high excitation energies the dication does not fully fragment until after tens to hundreds of picoseconds making it computationally too heavy to perform full MD simulations at the DFT level (see SI).

\subsubsection*{Potential energy surface calculations}
We have explored the PES using DFT, in particular we have employed the B3LYP functional\cite{Lee1988,Miehlich1989,Becke1993} in combination with the 6-31G(d) basis set. 
This part of the study provides useful energetic and structural information of the experimentally measured exit channels. In order to identify the most relevant stationary points of the PES we have adopted the following strategy:

1) We have first performed molecular dynamics simulations at the same DFT-B3LYP/6-31G(d) level, to mimic the evolution of the system during the first femtoseconds after the ionization and excitation. 
This part of the simulations also starts by computing the energy required to doubly ionize adamantane in a Franck–Condon transition from the optimized neutral structure.
This is our starting point for the dynamics. 160 trajectories were carried out using the ADMP method with a maximum propagation time of 500~fs, and considering a time step of $\Delta t=$ 0.1~fs, and a fictitious mass of $\mu=$0.1~au.
Thus, after propagation of the doubly charged excited adamantane, we have obtained the evolution of the system in the first femtoseconds.

2) Then, using the last step in the dynamics as an initial guess, we have optimized the geometry of the produced species.

3) Finally, geometry optimization of fragments observed in the experiments has also been computed. To this end, we have considered several structures for each ${\rm C}_n{\rm H}_x^{q+}$ fragment, thus obtaining the relative energy of the exit channels observed in the experiment.\\
Harmonic frequencies have been computed after geometry optimization to confirm that the obtained structures are actual minima in the PES (no imaginary frequencies) and to evaluate the Zero-Point-Energy correction.
These calculations were carried out with the Gaussian09 package\cite{Gaussian09}.
The proposed strategy was used in the past with success to study the fragmentation dynamics of ionized biomolecules (see e.g. \cite{Maclot2013,Piekarski2015,Maclot2016}).

\subsubsection*{Statistical fragmentation simulations}
We study statistical fragmentation of doubly ionized adamantane with the recently developed M3C method\cite{Aguirre2017,M3Csoftware}, using the constrained approach presented in \cite{Aguirre2019}.
The key aspect of this methodology is that it provides a random way to move in the phase space (the so-called Markov chain) until a region of maximum entropy is reached, where the physical observables are computed. This description should be equivalent to an MD simulation in the infinite integration time limit.
In this work, we focus on two observables: 1) the probability of each fragmentation channel as a function of internal energy (the so-called breakdown curves), and 2) the distribution of the internal energy of the system in its components.
This method was successfully used in the past to describe the fragmentation of carbonaceous species\cite{Martinet2004,DiazTendero2005,Aguirre2017,Erdmann2018,Aguirre2019}.

The main ingredients of the M3C simulations, i.e. structures, energies, and vibrational frequencies of the fragments, are those obtained in the PES exploration at the B3LYP/6-31G(d) level.
In total, 148 molecules are included in the fragmentation model (see SI for details). 
Geometries are available as an additional file in the SI and in the M3C-store project database~\cite{M3Cstore}.

The statistical simulations have been carried out such that the sum of angular momenta from all fragments exactly compensates the orbital momentum resulting in a total angular momentum equal to zero.
We set the radius of the system to 30.0 \AA.
Implementation of larger radii implies similar Coulomb interaction among fragments, but requires increased sampling to achieve convergence; on the other hand, a smaller radius results in an artificial overestimation of the fragments' angular momentum.
We have performed a scan of the internal energy from 0 to 10 eV. 10000 numerical experiments for each value of internal energy have been carried out in order to estimate the error in the computed observables (and thus to use them as convergence criteria).
The numerical experiments each differ from one another in their initial values for vibrational energy, angular momentum, and molecular orientation, which were randomly chosen. 
All numerical experiments start from the most stable structure of the doubly charged $\rm C_{10}H_{16}^{2+}$.
The sequence \texttt{5*V,T,R,S:0,5*V,T,R,S:-1:1} has been used as a Markov chain, including a total of 2000 events; among them 10\% have been used as a burn-in period (see ref.~\cite{Aguirre2017} for details). \\

In summary, a complete picture of the fragmentation of excited doubly-charged adamantane is obtained with the theoretical simulations: dynamic, energetic and entropic approaches allow us to infer the main factors governing the experimentally observed processes, also providing complementary information.

\section{Results and discussion}

\begin{figure*}
\includegraphics[width=1\textwidth]{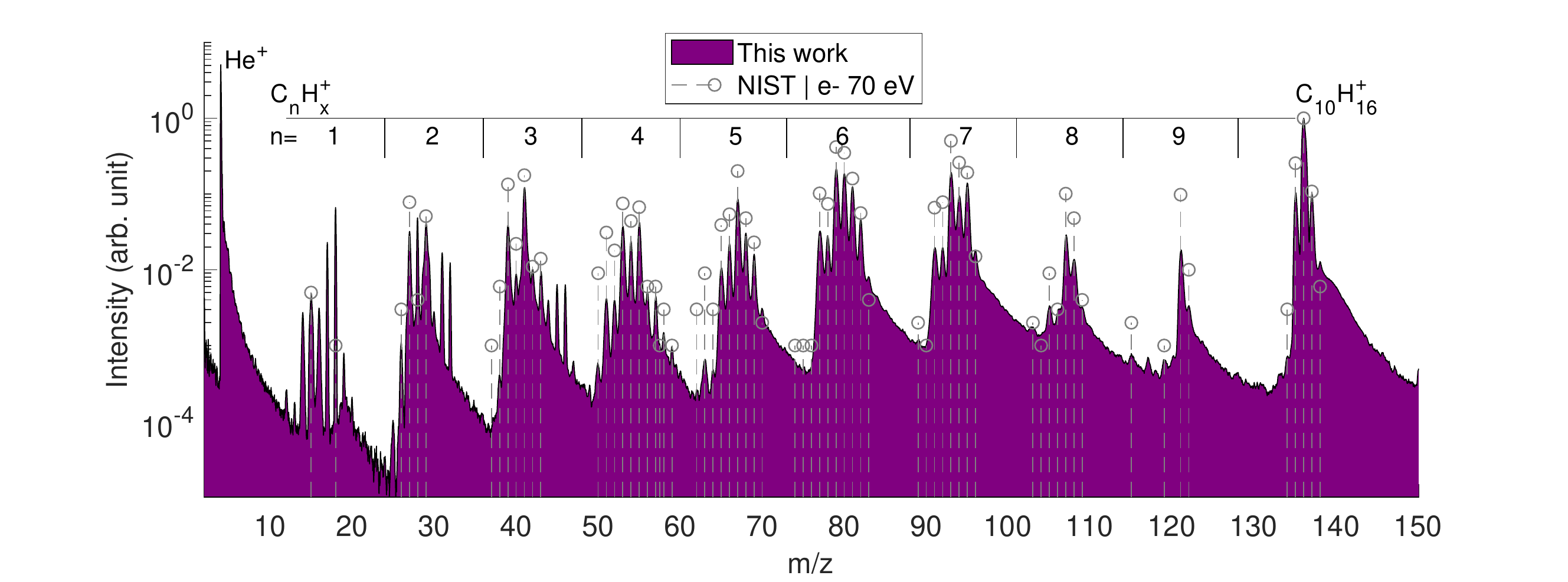}
\caption{\label{fig:MS} Mass spectrum of cationic products resulting from the interaction between XUV pulses and neutral adamantane molecules (purple). The dashed gray lines correspond to the mass spectrum obtained by electron impact at 70~eV\cite{Nist}.}
\end{figure*}

The experimental total mass spectrum of the charged products of interaction (Figure \ref{fig:MS}) is obtained by calibration of the TOF spectrum recorded after 275000 laser shots.
The most intense peak (excluding helium) corresponds to the singly charged parent ion at $m/z = 136$. The loss of one hydrogen atom is observed at a mass-to-charge ratio of $m/z = 135$ with an intensity of 9\% of the parent ion. 
The losses of two and three hydrogen atoms are also observed, however they have an intensity two orders of magnitude lower than that of the single hydrogen loss.
As a general feature, the production of a wide distribution of $\rm C_nH_x^+$ fragments resulting from dissociation of singly and doubly charged adamantane molecules is observed.
The most intense peaks of each $\rm C_n$ group are attributed to $\rm CH_3^+$, $\rm C_2H_5^+$, $\rm C_3H_5^+$, $\rm C_4H_7^+$, $\rm C_5H_7^+$, $\rm C_6H_7^+$, $\rm C_7H_9^+$, $\rm C_8H_{11}^+$ and $\rm C_9H_{13}^+$ (respectively  $m/z = 15$, 29, 41, 55, 67, 79, 93, 107 and 121) and are rather similar to the ones observed in the case of ionization by electron impact at 70~eV (dashed line in Figure~\ref{fig:MS}).
On the other hand, the main fragments of the $\rm C_n$ groups $n=3,4$ and 8 are different from the ones found by Candian \textit{et al.}\cite{Candian2018} (photodissociation around first ionization threshold), \textit{i.e.} $\rm C_3H_7^+$,$\rm C_4H_8^+$ and $\rm C_8H_{12}^+$ respectively, demonstrating that the dynamics of fragmentation is sensitive to the ionization/excitation energy.  

\begin{figure}
\centering
\includegraphics[width=.8\textwidth]{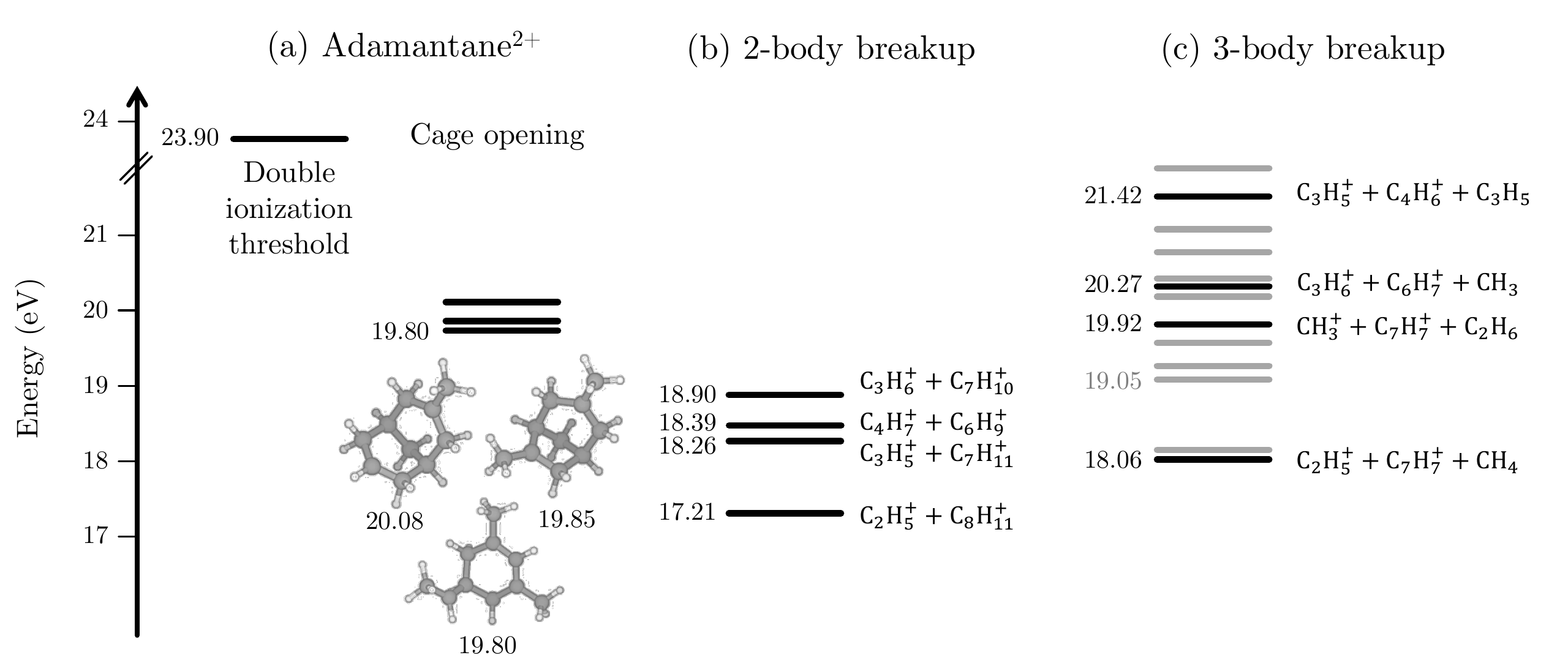}
\caption{\label{fig:PES} Key energy levels of adamantane dication processes. (a) Double ionization threshold and lowest energy configurations for doubly charged adamantane found in the PES exploration. (b) and (c), final energy levels of the fragmentation channels of the adamantane dication (2- and 3-body breakups) corresponding to the ones in Table~\ref{tab:frag}. Energy levels in gray are not explicitly labelled but can be found in Table~\ref{tab:frag}. The energy values are relative to the neutral ground state in units of eV.}
\end{figure}

Most of the fragments indicate strong intramolecular rearrangements with multiple hydrogen migrations and/or hydrogen losses. 
Some of these rearrangements may occur before fragmentation and lead to the cage opening of adamantane cations. The corresponding cage opening of the singly charged cation was already studied by Candian \textit{et al.} using DFT and RRKM simulations~\cite{Candian2018}. 
In the case of the doubly charged adamantane, the results of our {\it ab initio} molecular dynamics calculations (ADMP using DFT-B3LYP up to 500 fs) followed by PES exploration are summarized in Figure~\ref{fig:PES}. 
The three lowest energy configurations (Figure~\ref{fig:PES}(a)), appearing $\approx 4$~eV below the double ionization threshold, have an open-cage geometry and have at least one hydrogen migration ($\rm CH_3$ termination). 

\subsection*{Dication fragmentation pattern}
\begin{figure}
\centering
\includegraphics[width=.8\textwidth]{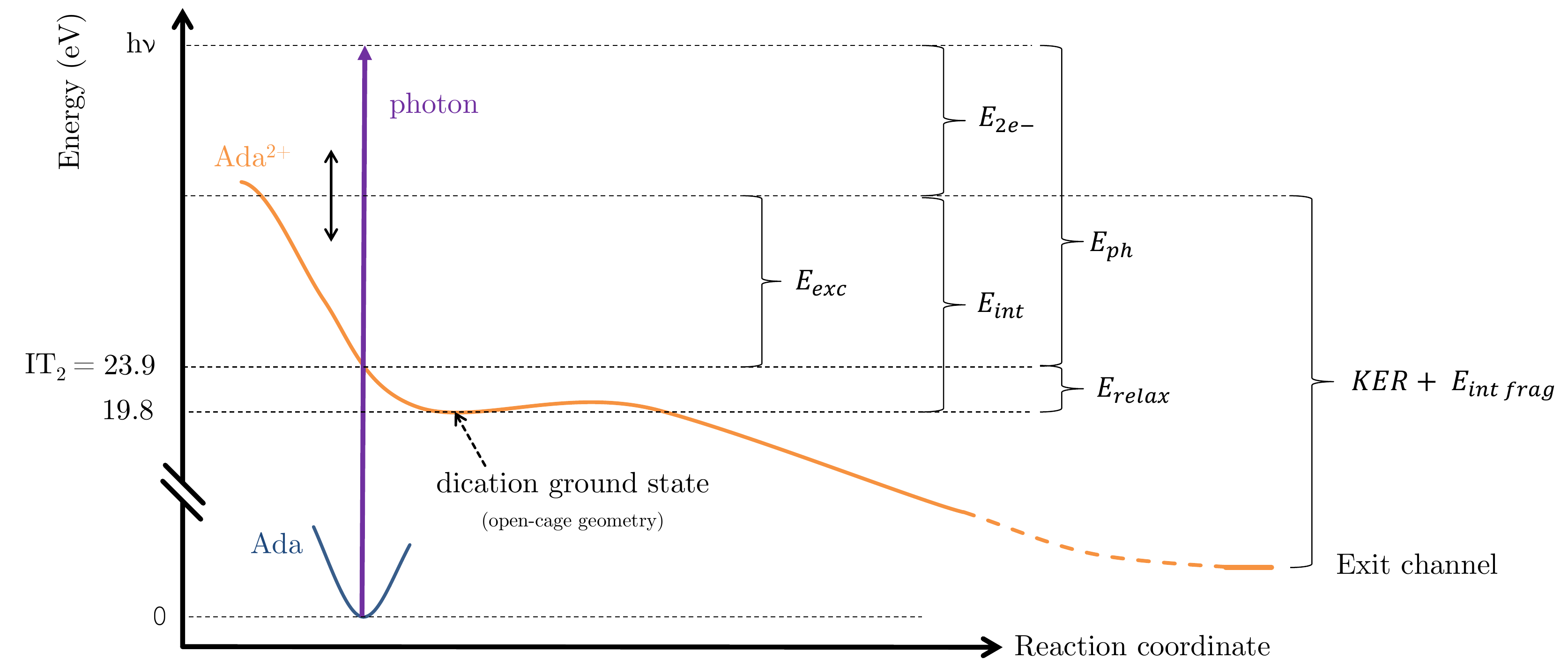}
\caption{\label{fig:PESscheme} Schematic of the dissociation of adamantane dication displaying the different energy quantities involved.}
\end{figure}
In order to help to understand the complex energetic picture of the dication dynamics, Figure~\ref{fig:PESscheme} shows a schematic dissociative potential energy curve for the dication, including the different energetic quantities that are useful for the discussion.
We consider the vertical ionization (Franck-Condon region) from the neutral ground state with a photon of energy $h \nu$ (purple arrow), such as $h \nu > \rm  IT_2$.
The excess energy after photoionization is defined as $\rm{E_{ph}} = \it{h \nu} - \rm IT_2$. 
We already know that the relaxation from the double ionization threshold $\rm IT_2$ to the ground state of the dication has a fixed energy of $\rm E_{relax} \approx 4.1$~eV (see Figure~\ref{fig:PES}(a)).
The kinetic energy of the electron pair involved in the double ionization is called $\rm E_{2e^-}$ and the internal energy of the dication ground state is denoted $\rm E_{int}$. 
Under our assumptions, the internal energy is the sum of the rapidly redistributed electronic excitation energy ($\rm E_{exc}$) and the relaxation energy ($E_{relax}$), and thus $\rm E_{int} \geq E_{relax} \approx 4.1$~eV. 
The final ionic products of the dissociation have a total energy equal to the sum of their internal energy ($\rm E_{int\,frag}$) and the kinetic energy release (KER). This total energy also corresponds to the sum of the initial internal energy of the dication ($\rm E_{int}$) and the energy difference between the dication ground state and the energy levels of the exit channel. 

\begin{figure}
\centering
\includegraphics[width=.9\textwidth]{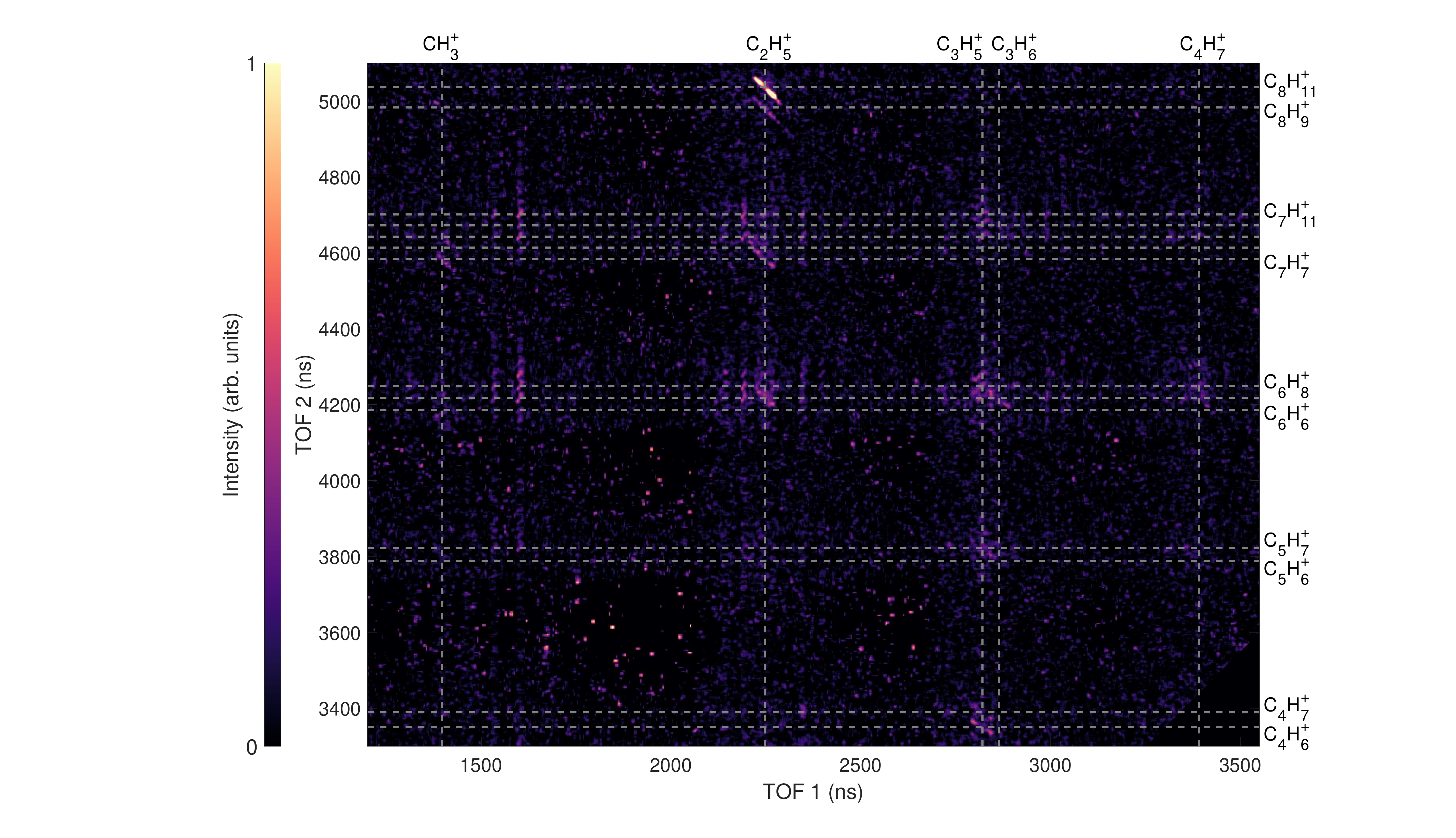}
\caption{\label{fig:Map} Ion-ion correlation map resulting from the fragmentation of adamantane dications.}
\end{figure}

Experimentally, the intact doubly charged parent ion is not observed in the mass spectrum (Figure \ref{fig:MS}) at the timescale of the detection (a few microseconds).
Moreover, no doubly charged fragments are detected. While the total mass spectrum is dominated by the fragments of singly charged adamantane molecules, the use of the partial covariance technique \cite{FRASINSKI1989,Frasinski2016} on the ion TOF spectrum enables to case correlate singly charged ions coming from the dissociation of the dication of adamantane using an ion-ion correlation map representation (Figure~~\ref{fig:Map}). 
Correlation islands in this map give the ion pairs that are summarized in Table~\ref{tab:frag} as well as the branching ratios (BR) of these fragmentation channels.
In addition, the PES exploration provides the final energy levels of the dication fragmentation channels referred to the neutral ground state ($\rm \Delta E$), which are represented in Figure~\ref{fig:PES} and given in Table~\ref{tab:frag}. 
The energy levels of the 2-body breakup channels appear at lower energies than most of the 3-body breakup ones. 
It is interesting to notice that all the energy levels are below the double ionization threshold, meaning that the dication of adamantane is metastable and will spontaneously dissociate (if no barrier, \textit{i.e.} transition state, at higher energy than the double ionization threshold is involved). 

\begin{table}
\centering
\caption{\label{tab:frag}List of correlated singly charged fragments coming from the dissociation of adamantane dication observed experimentally. In the case of $\rm n$-body breakups with $\rm n>2$, the neutral losses are given in mass losses such that the chemical formulae have to be seen as chemical element indicators and not necessarily as fragments. BR stands for branching ratio and is given in percent. $\rm \Delta E$ is the calculated final energy level of the dication fragmentation channels (2- and 3-body breakups) referred to the neutral ground state (in eV).}
\begin{tabular}{ccccccccc}
\multicolumn{2}{c}{Fragment 1}& \multicolumn{2}{c}{Fragment 2} & \multicolumn{2}{c}{Neutral loss} & BR & $\rm \Delta E$  \\
Formula & $\rm m$ (a.u.) & Formula & $\rm m$ (a.u.) & Formula & $\rm m$ (a.u.) & \% & (eV) \\ \hline
\multirow{4}{*}{CH$_3^+$} & \multirow{4}{*}{15} &  C$_8$H$_9^+$ & 105 & CH$_4$ & 16 & 2.8$\pm$ 0.2 & 19.27 & \\
& & C$_7$H$_9^+$ & 93 & C$_2$H$_4$ & 28 & 2.7$\pm$ 0.2 & 20.32   \\
& & C$_7$H$_7^+$ & 91 & C$_2$H$_6$ & 30 & 4.3$\pm$ 0.2 & 19.92  \\
& & C$_6$H$_6^+$ & 78 & C$_3$H$_7$ & 43 & 2.3$\pm$ 0.2 & 22.24 \\ \hline
\multirow{5}{*}{C$_2$H$_5^+$} & \multirow{5}{*}{29} & C$_8$H$_{11}^+$ & 107 & -- & -- & 23.3$\pm$ 0.3 & 17.21 \\
& & C$_8$H$_9^+$ & 105 & H$_2$ & 2 & 6.3$\pm$ 0.2 & 18.11  \\
& & C$_7$H$_8^+$ & 92 & CH$_3$ & 15 & 5.4$\pm$ 0.2 & 20.18  \\
& & C$_7$H$_7^+$ & 91 & CH$_4$ & 16 & 6.1$\pm$ 0.2 & 18.06   \\
& & C$_6$H$_7^+$ & 79 & C$_2$H$_4$ & 28 & 7.1$\pm$ 0.2 & 19.05 \\ \hline
\multirow{5}{*}{C$_3$H$_5^+$} & \multirow{5}{*}{41} & C$_7$H$_{11}^+$ & 95 & -- & -- & 2.5$\pm$ 0.2 & 18.26  \\
& & C$_6$H$_8^+$ & 80 & CH$_3$ & 15 & 5.3$\pm$ 0.2 & 20.31 \\
& & C$_5$H$_7^+$ & 67 & C$_2$H$_4$ & 28 & 5.3$\pm$ 0.2 & 19.78\\
& & C$_4$H$_7^+$ & 55 & C$_3$H$_4$ & 40 & 4.6$\pm$ 0.2 & 21.06 \\
& & C$_4$H$_6^+$ & 54 & C$_3$H$_5$ & 41 & 6.6$\pm$ 0.2 & 21.42 \\ \hline
\multirow{3}{*}{C$_3$H$_6^+$} & \multirow{3}{*}{42} & C$_7$H$_{10}^+$ & 94 & -- & -- & 2.8$\pm$ 0.2 & 18.90 \\
& & C$_6$H$_7^+$ & 79 & CH$_3$ & 15 & 6.4$\pm$ 0.2 & 20.27  \\
& & C$_5$H$_6^+$ & 66 & C$_2$H$_4$ & 28 & 3.7$\pm$ 0.2 & 20.80 \\ \hline
C$_4$H$_7^+$ & 55 & C$_6$H$_9^+$ & 81 & -- & -- & 2.3$\pm$ 0.2 & 18.39  \\
\end{tabular}
\end{table}

\subsection*{Statistical fragmentation simulations}

Assuming a low internal energy after the ionization, the fragmentation time of doubly charged adamantane is expected to be very long ($\gtrsim$100~ps, according to our DFTB simulation, see SI) and therefore we cannot afford to carry out simulations with {\it ab initio} molecular dynamics.
Thus, we use the M3C statistical method to obtain complementary information. This method was developed to study the fragmentation of molecular systems based on entropic criteria (see \cite{Aguirre2017,Aguirre2019,M3Csoftware} for details).
\subsubsection*{Breakdown curves}
\begin{figure}
\centering
\includegraphics[width=1\textwidth]{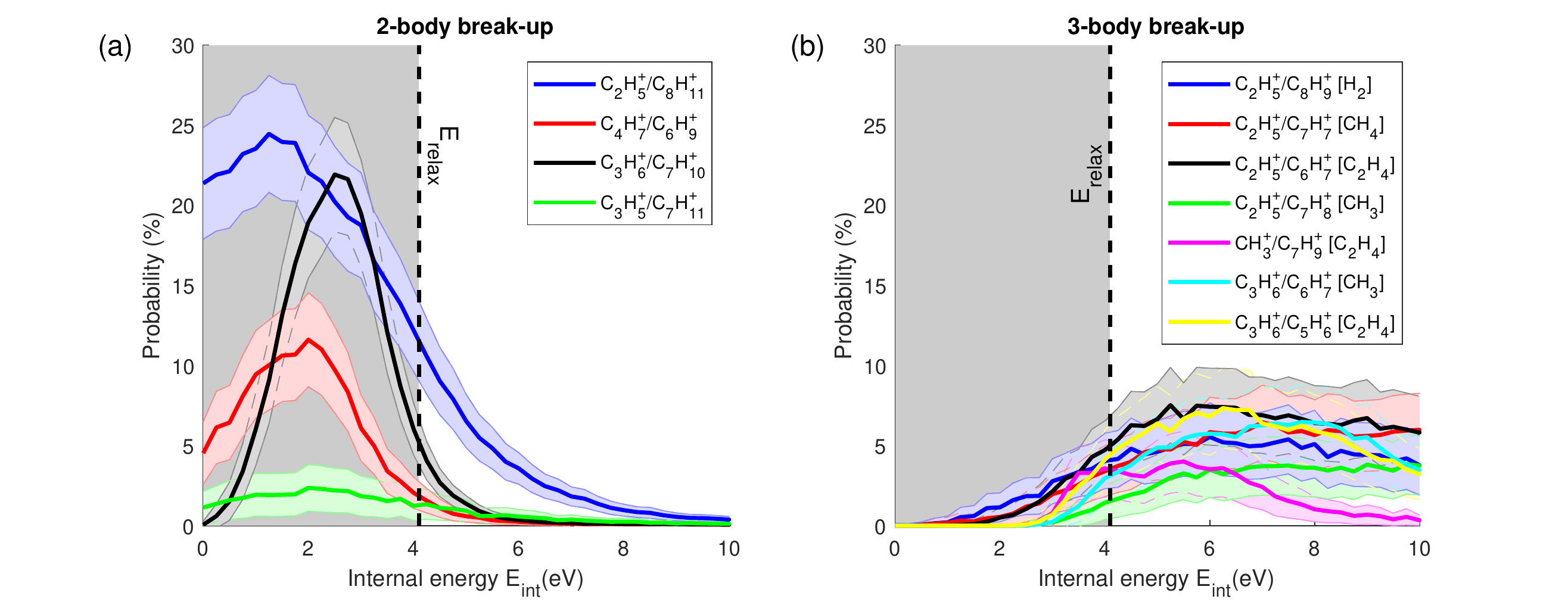}
\caption{\label{fig:M3C} Breakdown curves of the channels observed in the experiments for 2-body breakups (a) and 3-body breakups (b). The errors (shaded areas around curves) correspond to the standard deviation. The black dashed lines indicate the minimum of internal energy that we can reach in our case, meaning $\rm E_{relax}$, and the gray areas mark the regions of internal energy that are inaccessible in the experiment.}
\end{figure}

In Figure~\ref{fig:M3C}, showing the breakdown curves, the gray areas mark the inaccessible regions of internal energy ($\rm E_{int} < E_{relax}$). The probabilities for the 2-body channels (Figure~\ref{fig:M3C}(a)) all peak below $\rm E_{relax}$, with tails reaching into the accessible internal energy region. This is consistent with the fact that they appear in the lowest energy region in the PES (Figure~\ref{fig:PES}). Above $\rm E_{relax}$, only the $\rm C_2H_5^+/C_8H_{11}^+$ has a significant probability, while the other 2-body breakup channels $\rm C_3H_5^+/C_7H_{11}^+$, $\rm C_3H_6^+/C_7H_{10}^+$ and $\rm C_4H_7^+/C_6H_9^+$ are almost not populated. The 3-body breakups channels (Figure~\ref{fig:M3C}(b)) all peak around 6~eV, also consistent with their higher energy levels according to the PES in Figure~\ref{fig:PES}. 

Comparing the calculated breakdown cuves with the experimentally measured branching ratios in Table~\ref{tab:frag}, suggests that the internal energy of the adamantane dications is close to $\rm E_{relax}$ under the current conditions, since in this region the $\rm C_2H_5^+/C_8H_{11}^+$ channel dominates the other 2-body breakup channels, in good agreement with the experiment (BR$>$23\% vs. BR$<$3\%). If the internal energy was higher, the 3-body breakup channels would start to dominate, which is not observed in the experiment.

Under our assumption that the electronic excitation energy is considered to be rapidly redistributed into the nuclear degrees of freedom, this implies that most of the dications remain in the ground state or low excited states after ionization, and that the internal energy of the system mainly corresponds to the relaxation energy.

\subsubsection*{Energy storage in the fragments}
\begin{figure}
\centering
\includegraphics[width=0.7\textwidth]{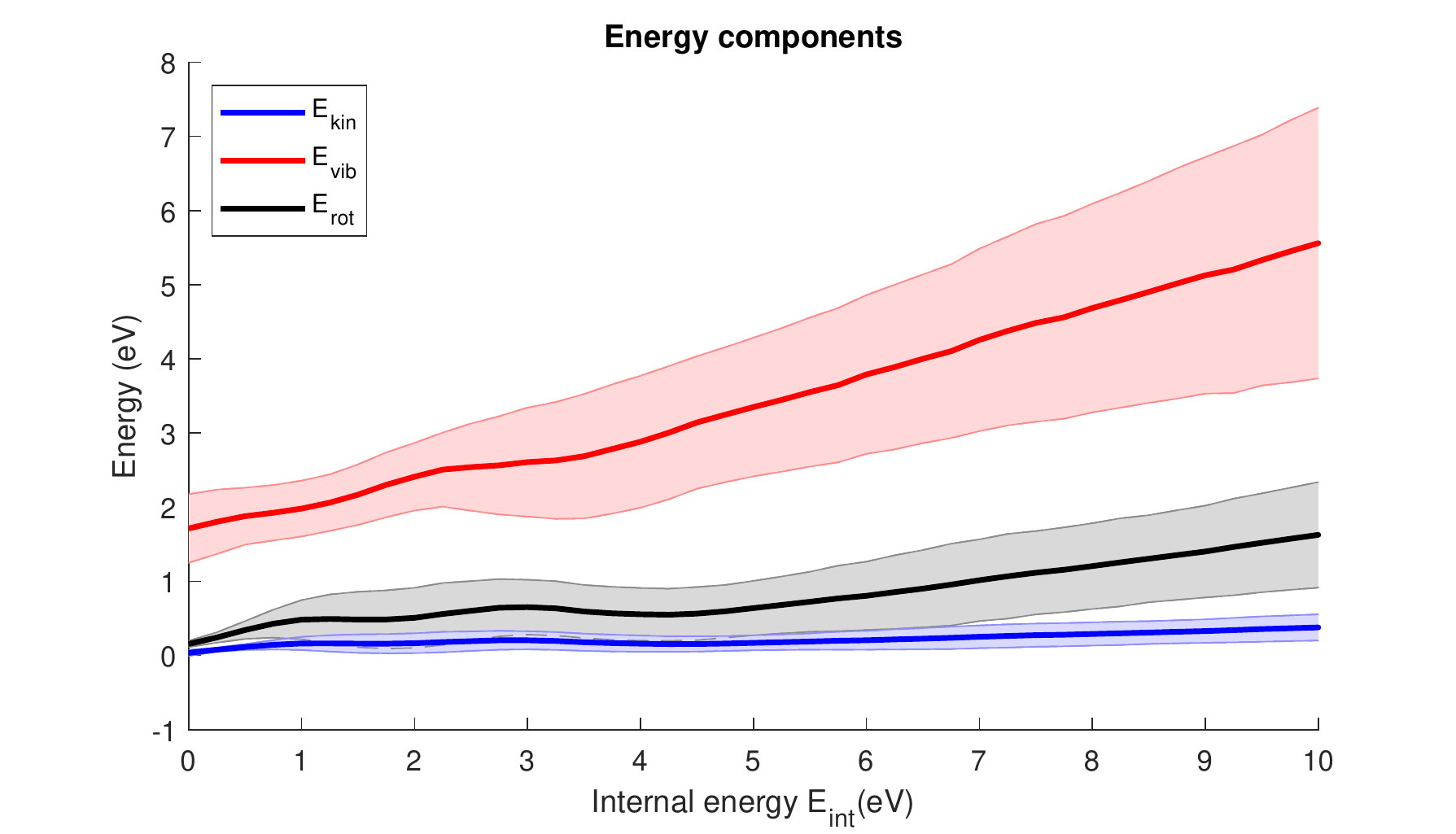}
\caption{\label{fig:M3C_NRJ} The average of the energy components $\rm E_\text{int}$, $\rm E_\text{vib}$, and $\rm E_\text{rot}$ is shown as a function of the internal energy. The errors (shaded areas around curves) correspond to the standard deviation.}
\end{figure}
We have further obtained additional valuable information by analyzing how the internal energy is distributed after ionization using the statistical simulations with the M3C code. 
Note that this analysis is based on the ergodic assumption, i.e. at infinite time when the system has reached equilibrium and the maximum entropy region in the phase space is populated. 

We can decompose the internal energy as $
\rm E_\text{int}=E_\text{kin}+E_\text{pot}+E_\text{vib}+E_\text{rot}$, where $\rm E_\text{kin}$, $\rm E_\text{pot}$, $\rm E_\text{vib}$, and $\rm E_\text{rot}$ represent the kinetic or translational, the potential, the vibrational, and the rotational energy components, respectively. The potential energy $\rm E_\text{pot}$ is the energy difference between different geometrical configurations, in this case the exit channel and the most stable dication structure (see Figure~\ref{fig:PES}).
While the potential energy is important for the total available energy, we now focus on how the latter is distributed between the remaining degrees of freedom, i.e. between $\rm E_\text{kin}$, $\rm E_\text{vib}$ and $\rm E_\text{rot}$. Figure~\ref{fig:M3C_NRJ} shows the average and the standard deviation of these energy components as a function of the internal energy $\rm E_\text{int}$. It is clear that in the considered internal energy range the vibrational contribution is most prominent, with smaller contributions of the rotational and kinetic energy components.
As already shown, the fragmentation is the dominant process in the relaxation of the adamantane dication; thus, the available energy of the system is primarily absorbed by the vibrational component, i.e. the produced fragments can store a large amount of energy in nuclear degrees of freedom (mainly vibrations). Assuming an internal energy of $\sim 4.1$~eV (corresponding to the relaxation energy), $\sim 70\%$ of the available energy is stored in vibration and $\sim 15\%$ in rotation while the remaining $15\%$ are shared among the other components. 

\subsection*{The main fragmentation channel C$_2$H$_{5}^+$/C$_{8}$H$_{11}^+$ - experimental evidence of a two-step process}
As we have seen, the channel C$_2$H$_{5}^+$/C$_{8}$H$_{11}^+$ is strongly dominating the fragmentation dynamics of the dication with a branching ratio $\sim23.3$\% whereas the other channels are at least three times less intense (Figure~\ref{fig:Map} and Table~\ref{tab:frag}).
This can be roughly interpreted by looking at the energy levels of the different fragmentation channels obtained from the exploration of the PES of the adamantane dication (Figure~\ref{fig:PES} and Table~\ref{tab:frag}). This main channel is energetically favorable since it has the lowest energy level at around $\sim$7~eV below the double ionization threshold. 
In addition, we have seen from the M3C calculations that the breakdown curve (Figure~\ref{fig:M3C}(a)) of this channel was dominant at low accessible internal energy ($\sim 4-5$~eV), thus being also entropically favorable in this energy region.

\begin{figure}
\centering
\includegraphics[width=.7\textwidth]{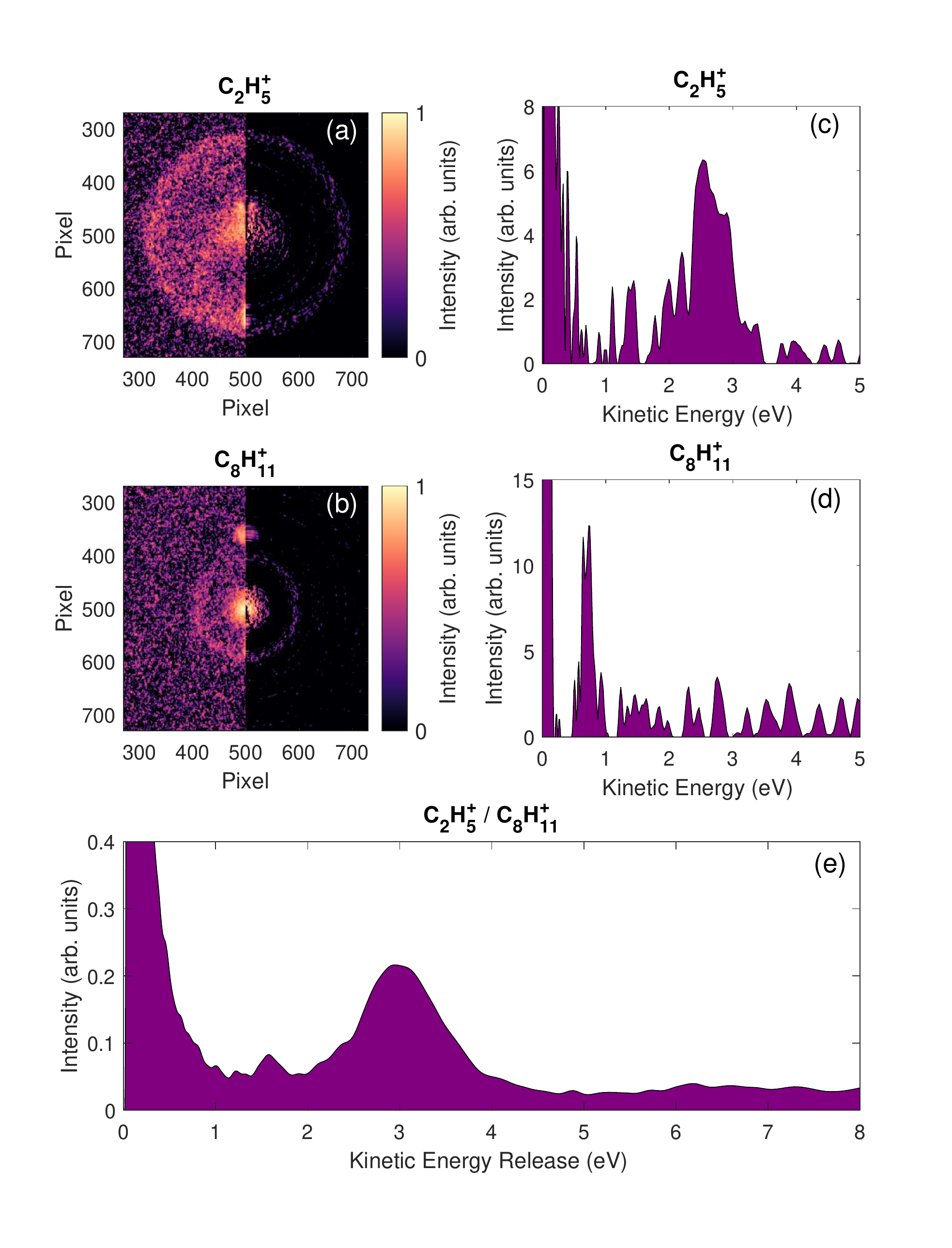}
\caption{\label{fig:KE} VMI images obtained after filtering using the partial covariance method on the TOF peaks correspond to the fragments C$_2$H$_5^+$ (a) and C$_8$H$_{11}^+$ (b) (left part: raw data and right part: inverted data). Artefacts are coming from intense signal (helium and parent ion) that were not filtered out by covariance analysis (see SI for more details). (c) and (d) Ion kinetic energy distributions of the respective fragments obtained by angular integration of the inverted data avoiding the artefacts signal and energy calibrated using ion trajectory simulations (SIMION\textregistered \cite{Dahl2000}). (e) Kinetic energy release distribution (KERd) for the channel C$_2$H$_5^+$ / C$_8$H$_{11}^+$ obtained by convolution of the kinetic energy distributions of the two fragments.}
\end{figure}

In order to have a better insight into the energetics of the main fragmentation channel, we can regard the ion kinematics of the dissociation process, particularly the kinetic energy release distribution (KERd).
Performing partial covariance analysis between the ion TOF and the ion VMI data gives the ''mass-selected'' velocity map images displayed on the left parts of panels (a) and (b) in Figure~\ref{fig:KE}. 
The right parts in Figure~\ref{fig:KE}(a) and (b) show the result of Abel inversion using an iterative method~\cite{Vrakking2001}. 
The angular integration of the inverted images gives, after energy calibration, the kinetic energy distributions of the fragments C$_2$H$_5^+$ and C$_8$H$_{11}^+$ (Figure~\ref{fig:KE}(c) and (d)).
The intense signal close to zero corresponds to the contribution from the dissociation of the singly charged adamantane.
At higher energies, clear peaks show contributions at $\sim$2.5~eV and $\sim$0.7~eV respectively due to the Coulomb repulsion of the 2-body breakup of the dication and verify the momentum conservation principle.
It is possible to obtain the KERd of the channel C$_2$H$_5^+$/C$_8$H$_{11}^+$ by convolution of the two individual kinetic energy distributions (Figure~\ref{fig:KE}(e)). The peak at $\sim$3~eV indicates the main energy contribution of this channel.

The small value of the KER reflects complex fragmentation dynamics with strong molecular rearrangement before the charge separation takes place: In a first step, a cage opening leading, most probably, to one of the structures in Figure~\ref{fig:PES}; then a second molecular reorganization producing both charged fragments; and finally charge repulsion between them.
The energy difference between the open structures in Figure~\ref{fig:PES} and the C$_2$H$_{5}^+$/C$_{8}$H$_{11}^+$ exit channel is between $\sim$2.6 and $\sim$2.9~eV, which is consistent within a multiple step fragmentation as the one presented here.
Considering the Coulomb repulsion between the two positive charges, the energy released is $C/R$, with $C=14.4$~eV$\rm \cdot \mathring{A}$ and $R$ the initial inter-charge distance in $\rm[\mathring{A}]$. An energy release of $\sim$~3.0~eV, as measured in the experiment, corresponds to an inter-charge distance of 4.8~$\rm \mathring{A}$.
However, the maximum distance between two C atoms in the closed-cage adamantane structure is $\sim~3.5~\rm \mathring{A}$, discarding the assumption of an instantaneous double ionization  followed by prompt fragmentation.
Thus, structural rearrangements before the charge separation would produce a considerable extension of the structure, thus increasing the inter-charge distance up to 4.8~$\rm \mathring{A}$ (as inferred from the experiment). 
This further confirms the multi-step processes with cage opening preceding Coulomb repulsion.

\subsection*{Photoelectron spectrum}
\begin{figure}[!b]
\centering
\includegraphics[width=.7\textwidth]{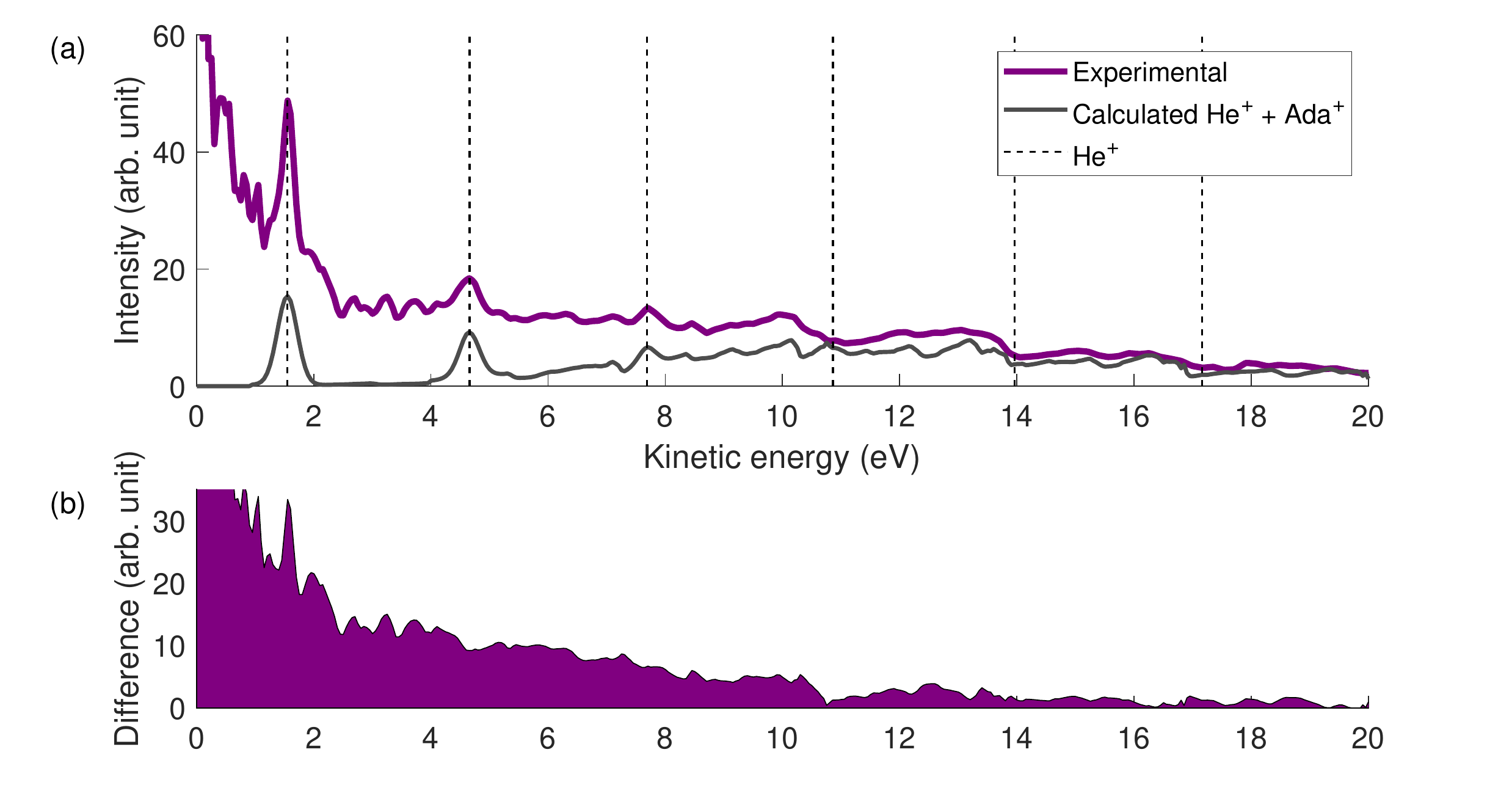}
\caption{\label{fig:Photoelec}(a) Total photoelectron spectrum (purple). The black dashed lines indicate the position of helium photoelectrons. The gray line is the calculated sum of helium photoelectrons (Gaussian) and of the photoelectrons coming from the singly charged adamantane (taken from~\cite{Schmidt1973}) considering our XUV spectrum and the photoabsorption cross section of adamantane~\cite{Steglich2011}. (b) Estimated photoelectron spectrum coming from double ionization of adamantane obtained as the difference between the purple and the gray curves of panel (a).}
\end{figure}

Figure~\ref{fig:Photoelec}(a) (purple line) shows the total photoelectron energy spectrum obtained by angular integration of the electron VMI data after inversion using an iterative method~\cite{Vrakking2001}.
Electrons coming from the ionization of the helium buffer gas can be seen around the dashed lines and were used for energy calibration. 
The contribution to the photoelectron spectrum associated with single ionization by the harmonics is found in the higher energy part of the total spectrum. 
The expected contribution, shown by the gray line Figure~\ref{fig:Photoelec}(a), is calculated using the measured photoelectron spectrum for singly charged adamantane~\cite{Schmidt1973}, the harmonic spectrum (see Figure~\ref{fig:XUVSpec}), and the photoabsorption cross section of adamantane~\cite{Steglich2011}. 
In this estimated contribution, the calculated photoelectron spectrum from ionization of helium has also been included, and by subtracting this from the total photoelectron spectrum, the expected spectrum of the photoelectrons associated with double ionization is obtained and shown in Figure~\ref{fig:Photoelec}(b).

Although the absolute signal in the resulting spectrum is sensitive to the scaling of the calculated spectrum, it is clear that the photoelectrons primarily occupy the low-energy part of the spectrum ($<10$~eV) compared to the photoelectrons from single ionization. From the harmonic spectrum (Figure~\ref{fig:XUVSpec}) the maximum total electron pair energy is $\rm E_{2e-} \approx 21$~eV, assuming that the dications remain in the ground state or in low excited states after ionization. Thus, the observed cut-off at $\sim 10$~eV, suggests that the available energy is shared rather evenly between the two electrons. All-in-all, the shape of the photoelectron energy distribution supports the assumption of low excitation, apart from the unexpectedly strong contribution at energies below 2~eV. The latter feature is a possible indication of excitation of higher-lying electronic states, resulting in low energy electrons. Possible candidates for such states are found through a calculation of the excited states of the adamantane dication (see SI for more details) exhibiting a dense band of excited states between 35 and 40~eV (relative to the ground state of the neutral) that would result in large internal energy values $\rm E_{int}$ between 15 and 20~eV, and values of $\rm E_{2e-}$ in the 5-10~eV range following ionization by the cut-off harmonics. Such excitation could not be inferred from the experimental fragmentation pattern, nor is included in the current level of theory, which calls for further investigation.

\section{Conclusion}
We have performed a detailed study of the photodissociation of adamantane, focused on the fragmentation dynamics of the dication. 
By combining the experimental analysis with multiple theoretical methods we unraveled key processes governing charge and energy distribution after the ultrafast photoionizaton and the subsequent fragmentation dynamics. 

We found that the most stable structures of the dication of adamantane present an open-cage geometry, appearing at $\sim 4$~eV below the double ionization threshold, that can be reached in a few tens of femtoseconds after the ionization. However, these structures are metastable and evolve producing several fragments in a Coulomb repulsion process.
Much like other carbonaceous species, adamantane dications are quite efficient energy reservoirs, being able to store a large amount of energy in particular in vibrational modes. This occurs when the internal energy generated in the photoionization, together with the energy produced in the Coulomb explosion, is redistributed among the nuclear degrees of freedom of the produced fragments.

Among the different fragmentation pathways, the most populated channel $\rm C_2H_5^+/C_8H_{11}^+$ is the lowest in energy in the PES ($\sim2.6$~eV below the most stable structure of the adamantane dication) leading us to conclude that the fragment distribution is mainly governed by energetic criteria. 
We have shown that the ion KERd of this channel peaks at $\sim 3$~eV with a width of $1$~eV allowing us to experimentally confirm that the cage opening takes place prior to fragmentation.
A qualitative comparison between the experimental branching ratios and the results of statistical fragmentation simulations suggests that the internal energy of the dication largely consists of the relaxation energy from the cage opening, with only minor contribution from redistribution of electronic excitation energy.
Finally, measurements of the photoelectron kinetic energy spectrum largely confirms these observations, but also indicate the possible existence of electronic excitation which could not be further investigated in the current experiment.

The presented results highlight the complexity of, and provides pieces of information on, the fragmentation of multiply charged diamondoids. While this study was able to identify the dominant fragmentation pathways and shine light on the redistribution of energy and charge, the further elucidation of the ultrafast excitation dynamics, and in particular the timescales involved, calls for time-resolved experiments. Such experiments can be envisaged using ultrashort single wavelength XUV pulses, e.g. from free electron lasers, in combination with multicoincidence ion-electron spectroscopy techniques. 

\bibliography{Manuscript_Maclot}

\section*{Acknowledgements}
A special thanks is given to A. Persson for his support for the high-power laser and to M. Gisselbrecht for critical reading of the manuscript.
This research was supported by the Swedish Research Council, the Swedish Foundation for Strategic Research and the Crafoord Foundation.
This project has received funding from the European Union’s Horizon 2020 research and innovation program under the Marie Sklodowska-Curie Grant Agreement No. 641789 MEDEA and under grant agreement no 654148 Laserlab-Europe.
The research was conducted in the framework of the International Associated Laboratory (LIA) Fragmentation DYNAmics of complex MOlecular systems - DYNAMO, funded by the Centre National de la Recherche Scientifique (CNRS).
The authors acknowledge the generous allocation of computer time at the Centro de Computaci\'{o}n Cient\'{i}fica at the Universidad Aut\'{o}noma de Madrid (CCC-UAM).
This work was partially supported by the project CTQ2016-76061-P of the Spanish Ministerio de Econom\'{i}a y Competitividad (MINECO).
Financial support from the MINECO through the ``Mar\'{i}a de Maeztu'' Program for Units of Excellence in R\&D (MDM-2014-0377) is also acknowledged.

\section*{Author contributions statement}
P.R, P.J and S.M. designed the experiments, S.M., J.L., J.P., H.W., P.R., P.R., S.I., F.B., H.C.-A., and B.A.H. performed the experiment, S.M. and J.L analysed the results. S.D.-T. performed the potential energy surface and the molecular dynamics calculations, and N.F.A. the M3C simulations. S.M., S.D.-T. and P.J wrote the original draft, S.D.-T. and N.F.A. wrote the theoretical sections and part of the discussion and all authors reviewed the manuscript. 

\section*{Competing interests}
The authors declare no competing interests.

\section*{Additional information}
Supplementary information is available for this paper at \href{https://www.XXXXsomewhereXXXXX.com}{link}

\end{document}